\documentclass[a4paper,aps,twocolumn,nofootinbib]{revtex4}
\RequirePackage[colorlinks,hyperindex]{hyperref}
\RequirePackage[english]{babel}
\RequirePackage[latin1]{inputenc}
\RequirePackage[T1]{fontenc}
\RequirePackage{mathrsfs}
\RequirePackage{amsmath}
\RequirePackage{amssymb}
\RequirePackage{amsbsy}
\RequirePackage{color}
\RequirePackage{bm}
\RequirePackage{slashed}
\hypersetup{colorlinks=true,breaklinks=true,urlcolor=blue,linkcolor=red}
\begin{document}
%%%%%%%%%%%%%%%%%%%%%%%%%%%%%%%%%%%%%%%%%%%%%%%%%%%%%%%%%%%%%%%%%%%%%%%%%%%%%%%%%%%%%%%%%%%%%%%%%%%
\title{\bf{Restrictions on torsion-spinor field theory}}
\author{Luca Fabbri$^{\ast}$, Manuel Tecchiolli$^{\dagger}$}
\affiliation{$^{\ast}$DIME Sez. Metodi e Modelli Matematici, Universit\`{a} di Genova,
Via all'Opera Pia 15, 16145 Genova ITALY\\
$^{\dagger}$Department of Physics, ETH Z\"{u}rich,
H\"{o}nggerberg, CH-8093 Z\"{u}rich, SWITZERLAND}
\date{\today}
%%%%%%%%%%%%%%%%%%%%%%%%%%%%%%%%%%%%%%%%%%%%%%%%%%%%%%%%%%%%%%%%%%%%%%%%%%%%%%%%%%%%%%%%%%%%%%%%%%%
\begin{abstract}
Torsion propagation and torsion-spin coupling are studied in the perspective of the Velo-Zwanziger method of analysis; specifically, we write the most extensive dynamics of the torsion tensor and the most exhaustive coupling that is permitted between torsion and spinors, and check the compatibility with constraints and hyperbolicity and causality of field equations: we find that some components of torsion and many terms of the torsion-spin interaction will be restricted away and as a consequence we will present the most general theory that is compatible with all restrictions.
\end{abstract}
%%%%%%%%%%%%%%%%%%%%%%%%%%%%%%%%%%%%%%%%%%%%%%%%%%%%%%%%%%%%%%%%%%%%%%%%%%%%%%%%%%%%%%%%%%%%%%%%%%%
\maketitle
%%%%%%%%%%%%%%%%%%%%%%%%%%%%%%%%%%%%%%%%%%%%%%%%%%%%%%%%%%%%%%%%%%%%%%%%%%%%%%%%%%%%%%%%%%%%%%%%%%%
\section{INTRODUCTION}
It is almost a century that Einsteinian gravity has been complemented with torsion \cite{C1,C2,C3}, and many decades that its importance for the coupling to the spin of spinors was recognized \cite{S,K} (see also \cite{Hehl:2007bn} for recent review/overview).

However, the Sciama-Kibble completion of Einsteinian gravitation is still mainly focused on the most straightforward generalization of Einstein gravity, where torsion is not allowed to propagate, while torsion is in general a physical field, and therefore it must have a propagation implemented in terms of second-order derivatives of the torsional terms included into the action. In parallel, also the torsional coupling to the spin of spinors is still mainly focused on the simplest form, and from the perspective of studying the torsion-spin interaction in its most general form then all torsion-spinor terms are to be included too.

Mathematical generality is not the only reason to have higher-order mass dimension terms in the action, and one physical argument is that such terms may be important in situations involving Lorentz symmetry violations \cite{Kostelecky:2007kx,Lehnert:2013jsa}.

The problem we now face is that, in principle, we shall find an infinity of such terms, unless some concepts would intervene to restrain this profusion, and a first-level solution is to notice that because the spinorial field equations have first-order derivatives in the spinor fields, then the inclusion of higher-than-first-order derivatives of spinor fields would make no sense, so that we ought restrict the inventory to products of first-order derivatives of spinor fields with torsion at most. In this way however we can still have mass-dimension $5$ terms, and we cannot remove them by using the criterion of renormalizability \cite{Kostelecky:2007kx,Lehnert:2013jsa}.

Even with only mass-dimension $5$ terms, it is still possible to include several terms, and therefore a second-level solution would be to assess which of these terms are acceptable within the Velo-Zwanziger analysis \cite{Velo:1969bt,Velo:1970ur}.

So in this paper, we will first of all make the inventory of terms that can be included in the action: this means all possible terms that are quadratic in the derivatives of torsion, beside the usual Dirac term containing first-order derivatives of spinor fields, as well as all the interaction terms given by the products of derivatives of spinor and torsion, but also products of spinors and squared torsion down to the usual products of spinors and torsion; once this is done, we will proceed by removing all those terms that do not cope with the restrictions that are imposed by the implementation of the Velo-Zwanziger method \cite{Velo:1969bt,Velo:1970ur}.

As we are going to witness, there will only be few terms remaining after that this analysis gets implemented.
%%%%%%%%%%%%%%%%%%%%%%%%%%%%%%%%%%%%%%%%%%%%%%%%%%%%%%%%%%%%%%%%%%%%%%%%%%%%%%%%%%%%%%%%%%%%%%%%%%%
\section{Torsion and its propagation}
To begin we introduce the torsion tensor in general.

The torsion tensor $Q_{\rho\mu\nu}$ is a tensor of order three antisymmetric in two of its indices: as a consequence, it is always possible to decompose it according to
\begin{eqnarray}
&Q_{\rho\mu\nu}\!=\!\frac{1}{3}(g_{\rho\mu}Q_{\nu}\!-\!g_{\rho\nu}Q_{\mu})
\!+\!\frac{1}{6}W^{\alpha}\varepsilon_{\alpha\rho\mu\nu}\!+\!T_{\rho\mu\nu}
\end{eqnarray}
where $Q_{\nu}\!=\!Q^{\rho}_{\phantom{\rho}\rho\nu}$ is the trace and $W^{\alpha}\!=\!Q_{\rho\mu\nu}\varepsilon^{\rho\mu\nu\alpha}$ is the completely antisymmetric part and where $T_{\rho\mu\nu}$ such that it is $T^{\rho}_{\phantom{\rho}\rho\nu}\!=\!0$ and 
$T_{\rho\mu\nu}\varepsilon^{\rho\mu\nu\alpha}\!=\!0$ is called non-completely antisymmetric irreducible part of torsion. Therefore the trace has $4$ components while the completely antisymmetric dual has $4$ components and as a consequence the non-completely antisymmetric irreducible part of torsion has the $16$ components that remain to account for a total of $24$ components that the torsion tensor has in general.

When these geometric parts are taken as propagating fields, the number of their components has to be reduced to the number of their degrees of freedom, and because an $s$ spin field possesses 
$2s\!+\!1$ degrees of freedom, then all remaining non-physical components must be suppressed by requiring a suitable number of conditions of symmetry and contractions as well as divergences of the field.

The trace and the completely antisymmetric dual are vectors, that is spin-$1$ fields, and thus from the $4$ components, we have to isolate $3$ degrees of freedom, which has to be done by imposing $1$ constraint, and this is done by requiring some conditions on contractions and divergences: because vector fields are not reducible then it is only in terms of the divergence that the constraint must be implemented, and this is why for vector fields the divergence of the vector is what constitutes the constraint.

On the other hand, the non-completely antisymmetric irreducible part $T_{\rho\mu\nu}$ is a rank-$3$ tensor, that is a spin-$3$ field, so from the initial $16$ components, we must isolate $7$ degrees of freedom, thus we need $9$ constraints: all the symmetries and the contractions have already been used, hence again we can only work with the divergences of the tensor. We have two cases: the divergence $\nabla_{\rho}T^{\rho\mu\nu}$ that, due to symmetry and contraction properties, results into an antisymmetric tensor, therefore giving $6$ constraints in total; and the divergence $\nabla_{\nu}T^{\rho\mu\nu}$ that, due to symmetry and contraction properties, results into a traceless tensor, hence giving $15$ constraints in total. This means that the constraint $\nabla_{\rho}T^{\rho\mu\nu}$ can not be enough, while the constraint $\nabla_{\nu}T^{\rho\mu\nu}$ will always be too stringent, and thus no combination of constraints would ever be appropriate.

The fact that the non-completely antisymmetric irreducible part $T_{\rho\mu\nu}$ cannot be properly defined as a propagating physical field can also be seen dynamically.

The dynamical properties are studied by isolating the leading-derivative term, that is taking free propagation.

The propagation of the completely antisymmetric part or its dual axial-vector $W^{\alpha}$ and its curl $(\partial W)_{\alpha\nu}$ has been studied in \cite{Fabbri:2014dxa}: there it was found that $W^{\alpha}$ has the same dynamical properties of an axial-vector massive field, and that is an axial-vector Proca field, quite generally indeed.

For the trace vector $Q^{\alpha}$ and its curl $(\partial Q)_{\alpha\nu}$ we should study the propagation now: because the dynamics cannot be distinguished by the Velo-Zwanziger analysis only on the bases of parity-evenness or parity-oddness, then it is not surprising that performing on $Q^{\alpha}$ the Velo-Zwanziger analysis we find that $Q^{\alpha}$ has the same dynamical properties of a vector massive field, and that is a vector Proca field, also quite generally. For a good analogy, one might think at $Q^{\alpha}$ as some sort of massive electrodynamic field.

The propagation of the non-completely antisymmetric irreducible part of torsion is trickier: the Lagrangian has to be formed with squares of derivatives of $T_{\rho\mu\nu}$ and, as a quick inventory of all indices combination reveals, there are only three possible scalar terms, given by
\begin{eqnarray}
\nonumber
&\mathscr{L}\!=\!\frac{1}{2}\nabla_{\alpha}T_{\mu\nu\sigma}\nabla^{\alpha}T^{\mu\nu\sigma}+\\
\nonumber
&+\frac{3}{2}A\nabla_{\alpha}T^{\alpha\mu\nu}\nabla^{\beta}T_{\beta\mu\nu}
\!+\!3B\nabla_{\alpha}T^{\mu\nu\alpha}\nabla^{\beta}T_{\mu\nu\beta}-\\
&-\frac{1}{2}M^{2}T_{\mu\nu\sigma}T^{\mu\nu\sigma}\!+\!\mathscr{L}_{\mathrm{matter}}
\end{eqnarray}
with a mass term and a source Lagrangian.

The total Lagrangian, upon variation with respect to the $T_{\rho\mu\nu}$ field, would give the field equations
\begin{eqnarray}
\nonumber
&\nabla^{2}T_{\mu\nu\sigma}\!+\!A(2\nabla_{\mu}\nabla^{\beta}T_{\beta \nu\sigma}-\\
\nonumber
&-\nabla_{\sigma}\nabla^{\rho}T_{\rho\mu\nu}
\!+\!\nabla_{\nu}\nabla^{\rho}T_{\rho\mu\sigma}-\\
\nonumber
&-\nabla^{\alpha}\nabla^{\beta}T_{\beta \alpha\sigma}g_{\mu\nu}
\!+\!\nabla^{\alpha}\nabla^{\beta}T_{\beta \alpha\nu}g_{\mu\sigma})+\\
\nonumber
&+B(3\nabla_{\nu}\nabla^{\beta}T_{\mu\beta\sigma}
\!-\!3\nabla_{\sigma}\nabla^{\beta}T_{\mu\beta\nu}-\\
\nonumber
&-\nabla^{\alpha}\nabla^{\beta}T_{\alpha\beta\sigma}g_{\mu\nu}
\!+\!\nabla^{\alpha}\nabla^{\beta}T_{\alpha\beta\nu}g_{\mu\sigma}+\\
\nonumber
&+\nabla_{\mu}\nabla^{\rho}T_{\rho\nu\sigma}
\!+\!\nabla_{\sigma}\nabla^{\rho}T_{\rho\mu\nu}-\\
&-\nabla_{\nu}\nabla^{\rho}T_{\rho\mu\sigma})\!+\!M^{2}T_{\mu\nu\sigma}\!=\!S_{\mu\nu\sigma}
\label{fe}
\end{eqnarray}
where $S_{\mu\nu\sigma}$ is the source tensor obtained from the source Lagrangian, and where all the symmetry and the contraction properties of $T_{\rho\mu\nu}$ are inherited by its field equations themselves. What this means is that the constraints due to the symmetries and the traces are automatically implemented and consequently we only need to think about those that come from the  covariant divergences.

Once again, there are two possible cases, of which the first is the divergence with respect to the first index, and because we are studying free propagation, where in particular also gravity is absent, all curvatures are zero and we can commute all covariant derivatives getting
\begin{eqnarray}
\nonumber
&\nabla^{2}\nabla_{\mu}T^{\mu\nu\sigma}(1\!+\!2A\!+\!B)+\\
\nonumber
&+B(\nabla_{\mu}\nabla_{\beta}\nabla^{\nu}T^{\mu\beta\sigma}
\!-\!\nabla_{\mu}\nabla_{\beta}\nabla^{\sigma}T^{\mu\beta\nu})+\\
&+M^{2}\nabla_{\mu}T^{\mu\nu\sigma}\!=\!\nabla_{\mu}S^{\mu\nu\sigma}
\end{eqnarray}
where there appear third-order derivatives that convert this constraint into a field equation: therefore we have to ask $B=0$ and $1\!+\!2A=0$ to hold. With these restrictions and taking the divergence with respect to either of the remaining indices we obtain that
\begin{eqnarray}
\nonumber
&\nabla^{2}\nabla_{\nu}T^{\mu\nu\sigma}
\!-\!\frac{1}{2}(\nabla^{\mu}\nabla_{\beta}\nabla_{\nu}T^{\beta \nu\sigma}
\!+\!\nabla^{\sigma}\nabla_{\rho}\nabla_{\nu}T^{\rho\nu\mu}+\\
&+\nabla^{2}\nabla_{\rho}T^{\rho\mu\sigma})
\!+\!M^{2}\nabla_{\nu}T^{\mu\nu\sigma}\!=\!\nabla_{\nu}S^{\mu\nu\sigma}
\end{eqnarray}
where again there appear third-order derivatives that convert this constraint into a field equation, although now the absence of free parameters leaves us without any freedom to adjust the coefficients. Therefore third-order derivatives will always remain in what should otherwise be constraints, and thus they can not be acceptable.

The fact that the non-completely antisymmetric irreducible part $T_{\rho\mu\nu}$ as a propagating physical field does not have a match between degrees of freedom and independent field equations is in line with what we found above.

As a consequence, we are forced to the conclusion that the non-completely antisymmetric irreducible part of the torsion tensor is not well defined as a physical field.

This conclusion also stands in line with the trend that emerges from the Velo-Zwanziger analysis \cite{Velo:1970ur}: the scalar field is always well defined, the spin-$1$ field starts to display consistency issues and as the spin goes higher consistency problems tend to increase. In this paper, Velo and Zwanziger find that whereas spin-$1$ fields are still rather manageable, spin-$2$ fields require a number of constraints to be arbitrarily implemented for good position.

The spin-$3$ field should be even more problematic, and what we have discussed here shows that it is indeed.

We would also like to add that in the case that is given by the non-completely antisymmetric irreducible part of torsion, which comes from the geometry, there naturally is less freedom for adjustment, and we look at this rigidity as the reason for all the propagation problems.
%%%%%%%%%%%%%%%%%%%%%%%%%%%%%%%%%%%%%%%%%%%%%%%%%%%%%%%%%%%%%%%%%%%%%%%%%%%%%%%%%%%%%%%%%%%%%%%%%%%
\section{Torsion with Spin and their interactions}
We have dismissed such a non-completely antisymmetric part of torsion as not well defined in its propagation and thus as not physical. The trace part $Q^{\alpha}$ and the dual of the completely antisymmetric part $W^{\alpha}$ will be the only fields we shall consider: their dynamics are given by the vector and axial-vector massive Proca field equations.

As for matter fields we will only consider the spinorial field $\psi$ (with $\overline{\psi}$ as conjugate) defined upon introduction of the Clifford algebra $\boldsymbol{\gamma}^{a}$ from which $\left[\boldsymbol{\gamma}^{a}\!, \!\boldsymbol{\gamma}^{b}\right]\!=\! 4\boldsymbol{\sigma}^{ab}$ and the implicit $2i\boldsymbol{\sigma}_{ab}\!=\!\varepsilon_{abcd}\boldsymbol{\pi} \boldsymbol{\sigma}^{cd}$ are the relations defining the generators $\boldsymbol{\sigma}^{ab}$ of the spinor group and the parity-odd matrix $\boldsymbol{\pi}$ (which is merely the matrix usually indicated as gamma with an index five but in a notation in which the useless index is not in display): writing $\boldsymbol{\nabla}_{\mu}\psi$ as the covariant derivative of spinors, we have that its dynamics is given by the Dirac spinor field Lagrangian as it is usual.

With $Q^{\alpha}$ and $W^{\alpha}$ as well as $\overline{\psi}$ and $\psi$ and all combinations of the $\boldsymbol{\gamma}^{a}$ matrices, we can now come up with all the possible torsion-spinor interaction terms: those involving the coupling of spinors to the axial-vector torsion $W^{\alpha}$ are found in \cite{Fabbri:2018qzy}; again, because such a list of terms is quite independent on the field being an axial-vector or a vector, one may expect that very similar terms would appear for the vector torsion $Q^{\alpha}$ as well. This is indeed what shall happen; however, there are also properties that depend on the parity of the fields, and so some additional terms with products of vector $Q^{\alpha}$ and axial-vector $W^{\alpha}$ must be expected too. As an additional remark, we specify that all throughout this work we are going to consider only a Lagrangian that displays parity invariance. After having done the inventory of all possible terms, we obtain that
\begin{eqnarray}
\nonumber
&\mathscr{L}\!=\!-\frac{1}{4}(\partial W)^{2}\!+\!\frac{1}{2}M^{2}_{W}W^{2}
\!-\!\frac{1}{4}(\partial Q)^{2}\!+\!\frac{1}{2}M^{2}_{Q}Q^{2}+\\
\nonumber
&+i\overline{\psi}\boldsymbol{\gamma}^{a}\boldsymbol{\nabla}_{a}\psi\!-\!m\overline{\psi}\psi-\\
\nonumber
&-X_{W}\overline{\psi}\boldsymbol{\gamma}^{\mu}\boldsymbol{\pi}\psi W_{\mu}
\!-\!X_{Q}\overline{\psi}\boldsymbol{\gamma}^{\mu}\psi Q_{\mu}-\\
\nonumber
&-\overline{\psi}\psi\left(B_{W}W^{2}\!+\!B_{Q}Q^{2}\right)+\\
\nonumber
&+S_{W\!Q}i\overline{\psi}\boldsymbol{\pi}\psi WQ
\!+\!A_{W\!Q}2\overline{\psi}\boldsymbol{\sigma}^{ij}\boldsymbol{\pi}\psi W_{[i}Q_{j]}+\\
\nonumber
&+R_{W}\overline{\psi}\boldsymbol{\pi}\boldsymbol{\sigma}^{\mu\nu}\psi (\partial W)_{\mu\nu}
\!+\!D_{W}i\overline{\psi}\boldsymbol{\pi}\psi\nabla_{\mu}W^{\mu}+\\
\nonumber
&+R_{Q}i\overline{\psi}\boldsymbol{\sigma}^{\mu\nu}\psi(\partial Q)_{\mu\nu}
\!+\!D_{Q}\overline{\psi}\psi\nabla_{\mu}Q^{\mu}+\\
\nonumber
&+Y_{W}
i(\overline{\psi}\boldsymbol{\pi}\boldsymbol{\sigma}^{\mu\nu}\boldsymbol{\nabla}_{\mu}\psi\!-\!\boldsymbol{\nabla}_{\mu}\overline{\psi}\boldsymbol{\pi}\boldsymbol{\sigma}^{\mu\nu}\psi)W_{\nu}+\\
\nonumber
&+Y_{W}'\frac{1}{2}(\overline{\psi}\boldsymbol{\pi}\boldsymbol{\nabla}_{\mu}\psi\!-\!
\!\boldsymbol{\nabla}_{\mu}\overline{\psi}\boldsymbol{\pi}\psi)W^{\mu}+\\
\nonumber
&+Y_{Q}
(\overline{\psi}\boldsymbol{\sigma}^{\mu\nu}\boldsymbol{\nabla}_{\mu}\psi\!-\!\boldsymbol{\nabla}_{\mu}\overline{\psi}\boldsymbol{\sigma}^{\mu\nu}\psi)Q_{\nu}+\\
&+Y_{Q}'\frac{i}{2}(\overline{\psi}\boldsymbol{\nabla}_{\mu}\psi\!-\!
\!\boldsymbol{\nabla}_{\mu}\overline{\psi}\psi)Q^{\mu}
\label{L}
\end{eqnarray}
with $3$ mass terms and a total of $14$ coupling constants.

Its variation would yield the field equations given by
\begin{eqnarray}
\nonumber
&i\boldsymbol{\gamma}^{a}\boldsymbol{\nabla}_{a}\psi
\!-\!\left(m\!+\!B_{W}W^{2}\!+\!B_{Q}Q^{2}\right)\psi-\\
\nonumber
&-2Y_{W}iW_{\nu}
\boldsymbol{\pi}\boldsymbol{\sigma}^{\nu\mu}\boldsymbol{\nabla}_{\mu}\psi
\!+\!Y_{W}'W^{\mu}\boldsymbol{\pi}\boldsymbol{\nabla}_{\mu}\psi-\\
\nonumber
&-2Y_{Q}Q_{\nu}\boldsymbol{\sigma}^{\nu\mu}\boldsymbol{\nabla}_{\mu}\psi
\!+\!Y_{Q}'iQ^{\mu}\boldsymbol{\nabla}_{\mu}\psi-\\
\nonumber
&-X_{W}W_{\mu}\boldsymbol{\gamma}^{\mu}\boldsymbol{\pi}\psi
\!-\!X_{Q}Q_{\mu}\boldsymbol{\gamma}^{\mu}\psi +\\
\nonumber
&+S_{W\!Q}iWQ\boldsymbol{\pi}\psi
\!+\!A_{W\!Q}2W_{[i}Q_{j]}\boldsymbol{\sigma}^{ij}\boldsymbol{\pi}\psi+\\
\nonumber
&+\left(R_{W}\!+\!\frac{i}{2}Y_{W}\right)(\partial W)_{\mu\nu}
\boldsymbol{\pi}\boldsymbol{\sigma}^{\mu\nu}\psi+\\
\nonumber
&+\left(D_{W}\!-\!\frac{i}{2}Y_{W}'\right)\nabla_{\mu}W^{\mu}
i\boldsymbol{\pi}\psi+\\
\nonumber
&+\left(R_{Q}\!-\!\frac{i}{2}Y_{Q}\right)(\partial Q)_{\mu\nu}
i\boldsymbol{\sigma}^{\mu\nu}\psi+\\
&+\left(D_{Q}\!+\!\frac{i}{2}Y_{Q}'\right)\nabla_{\mu}Q^{\mu}\psi\!=\!0
\label{spinor}
\end{eqnarray}
as the spinor field equations, together with
\begin{eqnarray}
\nonumber
&\nabla_{\mu}(\partial W)^{\mu\nu}\!+\!(M^{2}_{W}
\!-\!2B_{W}\overline{\psi}\psi)W^{\nu}
\!=\!X_{W}\overline{\psi}\boldsymbol{\gamma}^{\nu}\boldsymbol{\pi}\psi-\\
\nonumber
&-S_{W\!Q}i\overline{\psi}\boldsymbol{\pi}\psi Q^{\nu}\!+\!2A_{W\!Q}Q_{\mu}
2\overline{\psi}\boldsymbol{\sigma}^{\mu\nu}\boldsymbol{\pi}\psi+\\
\nonumber
&+2R_{W}\nabla_{\mu}(\overline{\psi}\boldsymbol{\pi}\boldsymbol{\sigma}^{\mu\nu}\psi)
\!+\!D_{W}\nabla^{\nu}(i\overline{\psi}\boldsymbol{\pi}\psi)-\\
\nonumber
&-Y_{W}
i(\overline{\psi}\boldsymbol{\pi}\boldsymbol{\sigma}^{\mu\nu}\boldsymbol{\nabla}_{\mu}\psi\!-\!\boldsymbol{\nabla}_{\mu}\overline{\psi}\boldsymbol{\pi}\boldsymbol{\sigma}^{\mu\nu}\psi)-\\
&-Y_{W}'\frac{1}{2}(\overline{\psi}\boldsymbol{\pi}\boldsymbol{\nabla}^{\nu}\psi\!-\!
\!\boldsymbol{\nabla}^{\nu}\overline{\psi}\boldsymbol{\pi}\psi)
\end{eqnarray}
and
\begin{eqnarray}
\nonumber
&\nabla_{\mu}(\partial Q)^{\mu\nu}\!+\!(M^{2}_{Q}
\!-\!2B_{Q}\overline{\psi}\psi)Q^{\nu}
\!=\!X_{Q}\overline{\psi}\boldsymbol{\gamma}^{\nu}\psi-\\
\nonumber
&-S_{W\!Q}i\overline{\psi}\boldsymbol{\pi}\psi W^{\nu}\!-\!2A_{W\!Q}W_{\mu}
2\overline{\psi}\boldsymbol{\sigma}^{\mu\nu}\boldsymbol{\pi}\psi+\\
\nonumber
&+2R_{Q}\nabla_{\mu}(i\overline{\psi}\boldsymbol{\sigma}^{\mu\nu}\psi)
\!+\!D_{Q}\nabla^{\nu}(\overline{\psi}\psi)-\\
\nonumber
&-Y_{Q}
(\overline{\psi}\boldsymbol{\sigma}^{\mu\nu}\boldsymbol{\nabla}_{\mu}\psi\!-\!\boldsymbol{\nabla}_{\mu}\overline{\psi}\boldsymbol{\sigma}^{\mu\nu}\psi)-\\
&-Y_{Q}'\frac{i}{2}(\overline{\psi}\boldsymbol{\nabla}^{\nu}\psi\!-\!
\!\boldsymbol{\nabla}^{\nu}\overline{\psi}\psi)
\end{eqnarray}
as axial-vector torsion and vector torsion field equations.

On these equations, it is now the moment to perform the Velo-Zwanziger analysis. The Velo-Zwanziger analysis is summarized as follows: 1. for a given field equation, consider only the leading-derivative terms, which are the terms with the highest-order derivatives and being those determining the propagation; 2. in what remains, make the replacement $i\boldsymbol{\nabla}_{\alpha}\!\rightarrow\!n_{\alpha}$ so to focus on the normal to the surfaces of the wave fronts; 3. after such replacement, the remaining algebraic equation will have in general the form 
$\boldsymbol{A}\psi\!=\!0$ and because this must be valid for generally non-vanishing spinor fields $\psi$ then one has to require that $\mathrm{det}\boldsymbol{A}\!=\!0$ hold: this last equation is called characteristic equation, and its solutions in terms of the components of $n_{\alpha}$ are such that $n_{0}$ must be real or else the original field equations will not be hyperbolic, and then $n_{\alpha}$ must also be space-like or else the original field equations will not have causal structure. For further details, we refer to the seminal papers \cite{Velo:1969bt,Velo:1970ur}, where the authors also describe a variety of computational techniques that are useful when straightforward calculations cannot be done. In reference \cite{Velo:1970ur} in particular, various examples are also provided.

To perform the Velo-Zwanziger analysis on the spinorial field equations, we consider (\ref{spinor}) with the highest-order derivatives alone obtaining
\begin{eqnarray}
\nonumber
&i\boldsymbol{\gamma}^{a}\boldsymbol{\nabla}_{a}\psi
\!-\!2Y_{W}iW_{\nu}\boldsymbol{\pi}\boldsymbol{\sigma}^{\nu\mu}\boldsymbol{\nabla}_{\mu}\psi
\!+\!Y_{W}'W^{\mu}\boldsymbol{\pi}\boldsymbol{\nabla}_{\mu}\psi-\\
&-2Y_{Q}Q_{\nu}\boldsymbol{\sigma}^{\nu\mu}\boldsymbol{\nabla}_{\mu}\psi
\!+\!Y_{Q}'iQ^{\mu}\boldsymbol{\nabla}_{\mu}\psi\!=\!0
\end{eqnarray}
and replacing $i\boldsymbol{\nabla}_{\alpha}\!\rightarrow\! n_{\alpha}$ we get 
\begin{eqnarray}
\nonumber
&(\boldsymbol{\gamma}^{a}n_{a}\!-\!2Y_{W}W_{\nu}\boldsymbol{\pi}\boldsymbol{\sigma}^{\nu\mu}n_{\mu}
\!-\!iY_{W}'W^{\mu}\boldsymbol{\pi}n_{\mu}+\\
&+2iY_{Q}Q_{\nu}\boldsymbol{\sigma}^{\nu\mu}n_{\mu}\!+\!Y_{Q}'Q^{\mu}n_{\mu})\psi\!=\!0
\end{eqnarray}
in the form $\boldsymbol{A}\psi\!=\!0$ so that $\mathrm{det}\boldsymbol{A}\!=\!0$ is given by
\begin{eqnarray}
\nonumber
&\mathrm{det}|\boldsymbol{\gamma}^{\mu}n_{\mu}
\!+\!2Y_{W}\boldsymbol{\pi}\boldsymbol{\sigma}^{\mu\nu}n_{\mu}W_{\nu}
\!-\!iY_{W}'\boldsymbol{\pi}n_{\mu}W^{\mu}-\\
&-2iY_{Q}\boldsymbol{\sigma}^{\mu\nu}n_{\mu}Q_{\nu}
\!+\!Y_{Q}'n_{\mu}Q^{\mu}|\!=\!0\label{spinorche}
\end{eqnarray}
and this is the characteristic equation: its solutions must have $n_{0}$ real to ensure hyperbolicity, and then $n_{\alpha}$ either space-like or light-like to ensure causality. We shall next proceed to the evaluation of the characteristic equation.

The explicit form of (\ref{spinorche}) can be computed straightforwardly and the result is the expression given by
\begin{eqnarray}
\nonumber
&|n^{2}|^{2}[(1\!+\!W^{2}|Y_{W}|^{2}\!+\!|Y_{Q}|^{2}Q^{2})^{2}+\\
\nonumber
&+4|Y_{Q}|^{2}|Y_{W}|^{2}(|QW|^{2}\!-\!Q^{2}W^{2})]+\\
\nonumber
&+n^{2}[-2(|Y_{Q}|^{2}\!+\!|Y_{Q}'|^{2})\cdot\\
\nonumber
&\cdot(1\!-\!W^{2}|Y_{W}|^{2}\!+\!|Y_{Q}|^{2}Q^{2})
|Q\!\cdot\!n|^{2}-\\
\nonumber
&-2(|Y_{W}|^{2}\!+\!|Y_{W}'|^{2})\cdot\\
\nonumber
&\cdot(1\!+\!W^{2}|Y_{W}|^{2}\!-\!|Y_{Q}|^{2}Q^{2})
|W\!\cdot\!n|^{2}+\\
\nonumber
&+8Y_{Q}Y_{W}(Y_{W}'Y_{Q}'
\!-\!Y_{Q}Y_{W})\cdot\\
\nonumber
&\cdot (Q\!\cdot\!n)\ (W\!\cdot\!n)\ (Q\!\cdot\!W)]+\\
\nonumber
&+[2(|Y_{Q}|^{2}|Y_{W}|^{2}
\!+\!|Y_{Q}'|^{2}|Y_{W}'|^{2}-\\
\nonumber
&-|Y_{Q}'|^{2}|Y_{W}|^{2}
\!-\!|Y_{Q}|^{2}|Y_{W}'|^{2}-\\
\nonumber
&-4Y_{Q}'Y_{W}Y_{Q}Y_{W}')
|W\!\cdot\!n|^{2}|Q\!\cdot\!n|^{2}+\\
\nonumber
&+(|Y_{Q}|^{2}\!+\!|Y_{Q}'|^{2})^{2}|Q\!\cdot\!n|^{4}+\\
&+(|Y_{W}|^{2}\!+\!|Y_{W}'|^{2})^{2}|W\!\cdot\!n|^{4}]\!=\!0
\label{che}
\end{eqnarray}
symmetric for the interchange of the two vector torsion fields: this is what has to be discussed in special cases.

Because (\ref{che}) is rather complicated, it may be difficult to find all situations in which hyperbolicity and causality are respected, and therefore we face the problem from the opposite angle, trying to assess what are the instances in which acausality or lack of hyperbolicity occur.

To this purpose, consider that we can always approximate torsion to be weak, and in this circumstance we can always approximate both vectors to be small compared to the unity, therefore getting
\begin{eqnarray}
\nonumber
&|n^{2}|^{2}
\!-\!2n^{2}[(|Y_{Q}|^{2}\!+\!|Y_{Q}'|^{2})|Q\!\cdot\!n|^{2}+\\
\nonumber
&+(|Y_{W}|^{2}\!+\!|Y_{W}'|^{2})|W\!\cdot\!n|^{2}]+\\
\nonumber
&+[2(|Y_{Q}|^{2}|Y_{W}|^{2}
\!+\!|Y_{Q}'|^{2}|Y_{W}'|^{2}-\\
\nonumber
&-|Y_{Q}'|^{2}|Y_{W}|^{2}
\!-\!|Y_{Q}|^{2}|Y_{W}'|^{2}-\\
\nonumber
&-4Y_{Q}'Y_{W}Y_{Q}Y_{W}')
|W\!\cdot\!n|^{2}|Q\!\cdot\!n|^{2}+\\
\nonumber
&+(|Y_{Q}|^{2}\!+\!|Y_{Q}'|^{2})^{2}|Q\!\cdot\!n|^{4}+\\
&+(|Y_{W}|^{2}\!+\!|Y_{W}'|^{2})^{2}|W\!\cdot\!n|^{4}]\!\approx\!0
\label{cheapprox}
\end{eqnarray}
which is in fact easier to manipulate. A similar working hypothesis is that of considering cases where the vector $Q$ is smaller than the axial vector $W$ so that we obtain
\begin{eqnarray}
\nonumber
&|n^{2}|^{2}
\!-\!2n^{2}(|Y_{W}|^{2}\!+\!|Y_{W}'|^{2})|W\!\cdot\!n|^{2}+\\
&+(|Y_{W}|^{2}\!+\!|Y_{W}'|^{2})^{2}|W\!\cdot\!n|^{4}\!\approx\!0
\end{eqnarray}
which admits the only solution
\begin{eqnarray}
&n^{2}\!\approx\!
(|Y_{W}|^{2}\!+\!|Y_{W}'|^{2})|W\!\cdot\!n|^{2}
\end{eqnarray}
for which $n^{2}$ will always be positive, the wave fronts will always be space-like, and the propagation will always be acausal; because for $Y_{W}\!\neq\!0$ or $\!Y_{W}'\!\neq\!0$ it is always possible to find situations where acausality arises, then to ensure causality we must have $Y_{W}\!=\!Y_{W}'\!=\!0$ identically valid as constraints. Then (\ref{cheapprox}) reduces to the simpler
\begin{eqnarray}
\nonumber
&|n^{2}|^{2}
\!-\!2n^{2}(|Y_{Q}|^{2}\!+\!|Y_{Q}'|^{2})|Q\!\cdot\!n|^{2}+\\
&+(|Y_{Q}|^{2}\!+\!|Y_{Q}'|^{2})^{2}|Q\!\cdot\!n|^{4}\!\approx\!0
\end{eqnarray}
which admits the only solution
\begin{eqnarray}
&n^{2}\!\approx\!
(|Y_{Q}|^{2}\!+\!|Y_{Q}'|^{2})|Q\!\cdot\!n|^{2}
\end{eqnarray}
for which $n^{2}$ will always be positive, the wave fronts will always be space-like, and the propagation will always be acausal; because for $Y_{Q}\!\neq\!0$ or $\!Y_{Q}'\!\neq\!0$ it is always possible to find situations where acausality arises, then to ensure causality we must have $Y_{Q}\!=\!Y_{Q}'\!=\!0$ holding identically.

As $Y_{W}\!=\!Y_{W}'\!=\!Y_{Q}\!=\!Y_{Q}'\!=\!0$ then (\ref{che}) reduces to $n^{2}\!=\!0$ so that wave fronts are light-like and therefore causality is ensured. Then since $n^{2}\!=\!0$ implies $|n^{0}|^{2}\!=\!\vec{n}\cdot\vec{n}$ we also have that $n^{0}$ is real and hyperbolicity is ensured as well.

In studying under what circumstances (\ref{che}) may have unacceptable solutions, we have also obtained under what circumstances acceptable solutions are ensured.

So, because we have found that acausality can always occur unless $Y_{W}\!=\!Y_{Q}\!=\!Y_{W}'\!=\!Y_{Q}'\!=\!0$ identically, then we can conclude our analysis by saying that the four constraints given by $Y_{W}\!=\!Y_{Q}\!=\!Y_{W}'\!=\!Y_{Q}'\!=\!0$ are a necessary condition for causality, but also that they are a sufficient condition for causality and additionally they are a sufficient condition for hyperbolicity of field equations (\ref{spinor}).

The spinor field equations (\ref{spinor}) thus reduce to
\begin{eqnarray}
\nonumber
&i\boldsymbol{\gamma}^{a}\boldsymbol{\nabla}_{a}\psi
\!-\!\left(m\!+\!B_{W}W^{2}\!+\!B_{Q}Q^{2}\right)\psi-\\
\nonumber
&-X_{W}W_{\mu}\boldsymbol{\gamma}^{\mu}\boldsymbol{\pi}\psi
\!-\!X_{Q}Q_{\mu}\boldsymbol{\gamma}^{\mu}\psi +\\
\nonumber
&+S_{W\!Q}iWQ\boldsymbol{\pi}\psi
\!+\!A_{W\!Q}2W_{[i}Q_{j]}\boldsymbol{\sigma}^{ij}\boldsymbol{\pi}\psi +\\
\nonumber
&+R_{W}(\partial W)_{\mu\nu}\boldsymbol{\pi}\boldsymbol{\sigma}^{\mu\nu}\psi
\!+\!D_{W}\nabla_{\mu}W^{\mu}i\boldsymbol{\pi}\psi+\\
&+R_{Q}(\partial Q)_{\mu\nu}i\boldsymbol{\sigma}^{\mu\nu}\psi
\!+\!D_{Q}\nabla_{\mu}Q^{\mu}\psi\!=\!0
\end{eqnarray}
while the torsion axial-vector field equations are
\begin{eqnarray}
\nonumber
&\nabla_{\mu}(\partial W)^{\mu\nu}\!+\!(M^{2}_{W}
\!-\!2B_{W}\overline{\psi}\psi)W^{\nu}
\!=\!X_{W}\overline{\psi}\boldsymbol{\gamma}^{\nu}\boldsymbol{\pi}\psi-\\
\nonumber
&-S_{W\!Q}i\overline{\psi}\boldsymbol{\pi}\psi Q^{\nu}\!+\!2A_{W\!Q}Q_{\mu}
2\overline{\psi}\boldsymbol{\sigma}^{\mu\nu}\boldsymbol{\pi}\psi+\\
&+2R_{W}\nabla_{\mu}(\overline{\psi}\boldsymbol{\pi}\boldsymbol{\sigma}^{\mu\nu}\psi)
\!+\!D_{W}\nabla^{\nu}(i\overline{\psi}\boldsymbol{\pi}\psi)\label{W}
\end{eqnarray}
and the torsion vector field equations are
\begin{eqnarray}
\nonumber
&\nabla_{\mu}(\partial Q)^{\mu\nu}\!+\!(M^{2}_{Q}
\!-\!2B_{Q}\overline{\psi}\psi)Q^{\nu}
\!=\!X_{Q}\overline{\psi}\boldsymbol{\gamma}^{\nu}\psi-\\
\nonumber
&-S_{W\!Q}i\overline{\psi}\boldsymbol{\pi}\psi W^{\nu}\!-\!2A_{W\!Q}W_{\mu}
2\overline{\psi}\boldsymbol{\sigma}^{\mu\nu}\boldsymbol{\pi}\psi+\\
&+2R_{Q}\nabla_{\mu}(i\overline{\psi}\boldsymbol{\sigma}^{\mu\nu}\psi)
\!+\!D_{Q}\nabla^{\nu}(\overline{\psi}\psi)\label{Q}
\end{eqnarray}
and so far as the Velo-Zwanziger analysis is concerned we have that the field equations cannot be more general.
%%%%%%%%%%%%%%%%%%%%%%%%%%%%%%%%%%%%%%%%%%%%%%%%%%%%%%%%%%%%%%%%%%%%%%%%%%%%%%%%%%%%%%%%%%%%%%%%%%%
\section{Constant torsion}
Up to now we have seen how the Velo-Zwanziger analysis restricts the structure of the field equations. However, there are specific situations in which further reductions are implementable: in fact, one of the physical situations in which these field equations can be used is in the study of possible Lorentz symmetry violations \cite{Kostelecky:2007kx,Lehnert:2013jsa}.

In these papers, the authors consider Lagrangians such as the one we have examined here, and they show that a constant torsion may entail the break-down of a Lorentz symmetry. However, constant torsion cannot be assumed but must be obtained as solution of the torsion field equations in general: in \cite{Kostelecky:2007kx,Lehnert:2013jsa} the torsional field equations are never studied, and in fact they are not even presented.

In the present paper we have such field equations and therefore this study can be performed. A constant torsion is compatible with the torsion field equations whenever
\begin{eqnarray}
\nonumber
&(M^{2}_{W}
\!-\!2B_{W}\overline{\psi}\psi)W^{\nu}
\!=\!X_{W}\overline{\psi}\boldsymbol{\gamma}^{\nu}\boldsymbol{\pi}\psi-\\
\nonumber
&-S_{W\!Q}i\overline{\psi}\boldsymbol{\pi}\psi Q^{\nu}
\!+\!2A_{W\!Q}Q_{\mu}
2\overline{\psi}\boldsymbol{\sigma}^{\mu\nu}\boldsymbol{\pi}\psi+\\
&+2R_{W}\nabla_{\mu}(\overline{\psi}\boldsymbol{\pi}\boldsymbol{\sigma}^{\mu\nu}\psi)
\!+\!D_{W}\nabla^{\nu}(i\overline{\psi}\boldsymbol{\pi}\psi)
\label{constW}
\end{eqnarray}
and
\begin{eqnarray}
\nonumber
&(M^{2}_{Q}
\!-\!2B_{Q}\overline{\psi}\psi)Q^{\nu}
\!=\!X_{Q}\overline{\psi}\boldsymbol{\gamma}^{\nu}\psi-\\
\nonumber
&-S_{W\!Q}i\overline{\psi}\boldsymbol{\pi}\psi W^{\nu}
\!-\!2A_{W\!Q}W_{\mu}
2\overline{\psi}\boldsymbol{\sigma}^{\mu\nu}\boldsymbol{\pi}\psi+\\
&+2R_{Q}\nabla_{\mu}(i\overline{\psi}\boldsymbol{\sigma}^{\mu\nu}\psi)
\!+\!D_{Q}\nabla^{\nu}(\overline{\psi}\psi)
\label{constQ}
\end{eqnarray}
admit a source distribution for which they hold. But such an analysis is not necessary, as (\ref{constW}, \ref{constQ}) convert (\ref{L}) into
\begin{eqnarray}
\nonumber
&\mathscr{L}\!=\!i\overline{\psi}\boldsymbol{\gamma}^{a}\boldsymbol{\nabla}_{a}\psi
\!-\!m\overline{\psi}\psi-\\
&-\frac{1}{2}X_{W}\overline{\psi}\boldsymbol{\gamma}^{\mu}\boldsymbol{\pi}\psi W_{\mu}
\!-\!\frac{1}{2}X_{Q}\overline{\psi}\boldsymbol{\gamma}^{\mu}\psi Q_{\mu}
\end{eqnarray}
which is the Lagrangian we would have had in standard background with no higher-order mass dimension terms.

So if the Lagrangian considered in \cite{Kostelecky:2007kx} were to be taken in deep examination, it would become evident that there would be no Lorentz symmetry violation. Or at least that there would be no Lorentz symmetry violation apart from the one that would also be present in the standard case.

We conclude therefore that the assumption of constant torsion exceeds the boundary of its applicability.
%%%%%%%%%%%%%%%%%%%%%%%%%%%%%%%%%%%%%%%%%%%%%%%%%%%%%%%%%%%%%%%%%%%%%%%%%%%%%%%%%%%%%%%%%%%%%%%%%%%
\section{CONCLUSION}
In this paper, we have considered the Lagrangian that would arise from allowing all propagating torsional terms as well as all consistent interactions between torsion and spinor fields; after writing the Lagrangian, we proceeded in finding the field equations, studying them in terms of the Velo-Zwanziger method, removing all the terms that are found to be inconsistent with such an analysis.

As a first result, we found that, of the three irreducible parts of torsion, which we indicated with $T$, $Q$ and $W$, a number of restrictions took place for the non-completely antisymmetric irreducible component: we could not write field equations compatibly with the requirement that the number of degrees of freedom must match the number of independent field equations. This circumstance is consistent with the consideration that the higher the spin of a field the more difficult is to define its field equations.

Having then ruled out the non-completely antisymmetric irreducible component $T$ we have been left with the two vector components $Q$ and $W$ and as a consequence we built the Lagrangian with these two alone coupling to the spinor field: after obtaining the field equations, we computed the characteristic equations witnessing that it was always possible to find situations where acausal propagation would arise unless four coefficients were zero identically, and in doing so we established that the vanishing of these four coefficients is not only a necessary but also a sufficient condition for causality as well as for hyperbolicity, and we gave the most general field equations compatible with all restrictions of the Velo-Zwanziger analysis.

We finally considered works such as those of Kostelecky and co-workers about Lorentz symmetry violations based on the assumption of constant torsion, proving that when such models are studied in detail it is clear that the assumption of constant torsion would not give any Lorentz symmetry violation more than what would be in standard cases with no higher-order mass dimension terms.

In the extended Lagrangian (\ref{L}), the only non-restricted terms that can appear are summarized as follows: terms proportional to the constants $B_{W}$ and $B_{Q}$ can be seen as corrections to the masses of torsion and the mass of the spinor field; terms proportional to $R_{W}$ and $R_{Q}$ are those related to the coupling of torsion to the electric-like and magnetic-like di-pole spinor quantities; $D_{W}$ and $D_{Q}$ are the scalar analogous of the two we just mentioned, with coupling to the scalar and pseudo-scalar spinor quantities that would make them a mass and a pseudo-mass for the spinor field; terms in $S_{W\!Q}$ and $A_{W\!Q}$ instead do not have a clear interpretation because they describe a new type of interaction involving both vectors and the spinor field.

Just the same, the term in $S_{W\!Q}$ has a structure analogous to that of the term in $D_{W}$ while the term in $A_{W\!Q}$ has a structure analogous to that of the term in $R_{W}$ and as such, they represent some type of electric-like di-pole spinor quantity and some mass term of the spinor field.

The investigation of the physical effects of these additional terms, especially the properties of renormalizability of the Lagrangian, might be done in later works.

In order to study the cleanest background, one of our working hypothesis consisted in taking into account flat space-times, so a first opportunity for enlargement should come from considering the gravitational field effects.

Other opportunities for enlarging this analysis might come from relaxing any of the various working hypotheses we have done above, one of which being parity-evenness of the Lagrangian, so that a possible extension might be allowing parity-odd terms within the action \cite{Ho:2011xf}.

An extension of more general character is one involving the enlargement of the background, allowing the so-called non-metricity, beside torsion and curvature \cite{Hehl:1994ue}.

In this paper we have not considered such extensions because we are proceeding for increasing levels of complexity and in this perspective our aim was to study the most general decomposition of torsion, but the inclusion of non-metricity as final extension of the geometry, and parity-oddness as final extension of the dynamics, clearly are the next steps that should be done in this direction.

As has happened here, it may happen in either or both of these extensions that renormalizability issues arise, so that it is interesting to study how they may be linked to asymptotic safety, as it has been discussed in \cite{Mielke:2013vma}.

We again recall that for the Velo-Zwanziger analysis it is necessary to start with well-posed field equations.

For this reason it is essential that in all previously discussed extensions an investigation of the Cauchy problem be conducted as a preliminary study \cite{Hecht:1991jh}.
%%%%%%%%%%%%%%%%%%%%%%%%%%%%%%%%%%%%%%%%%%%%%%%%%%%%%%%%%%%%%%%%%%%%%%%%%%%%%%%%%%%%%%%%%%%%%%%%%%%
%%%%%%%%%%%%%%%%%%%%%%%%%%%%%%%%%%%%%%%%%%%%%%%%%%%%%%%%%%%%%%%%%%%%%%%%%%%%%%%%%%%%%%%%%%%%%%%%%%%

%%%%%%%%%%%%%%%%%%%%%%%%%%%%%%%%%%%%%%%%%%%%%%%%%%%%%%%%%%%%%%%%%%%%%%%%%%%%%%%%%%%%%%%%%%%%%%%%%%%
\end{document}